\documentclass[12pt]{article}

\usepackage{latexsym}
\usepackage{calc}
\usepackage{datetime}
\usepackage{amssymb}
\usepackage{graphicx}

\newcounter{hours}\newcounter{minutes}

\def\nr{\par \noindent}

\def\Def{\stackrel{\mathrm{def}}{=}}

\def\beq{\begin{equation}}
\def\eeq{\end{equation}}

\def\Z{\mathbb{Z}}

\def\BI{\begin{itemize}}
\def\EI{\end{itemize}}
\def\II{\item}
\newcommand{\SetEQ}{\setcounter{equation}{0}}
\newcommand{\refLE}[1]{\ensuremath{\stackrel{(\ref{#1})}{\leq}}}
\newcommand{\refEQ}[1]{\ensuremath{\stackrel{(\ref{#1})}{=}}}
\newcommand{\refGE}[1]{\ensuremath{\stackrel{(\ref{#1})}{\geq}}}

\newtheorem{theorem}{Theorem}
\newtheorem{lemma}{Lemma}
\newtheorem{corollary}{Corollary}

\newtheorem{assumption}{Assumption}
\newtheorem{axiom}{Axiom}
\newtheorem{definition}{Definition}

\newtheorem{example}{Example}
\newtheorem{remark}{Remark}
\newcommand{\proof}{\bf Proof: \rm \nr}
\newcommand{\qed}{\hfill $\Box$ \nr \medskip}

\def\ba{\begin{array}}
\def\ea{\end{array}}
\def\beann{\begin{eqnarray*}}
\def\eeann{\end{eqnarray*}}
\def\bea{\begin{eqnarray}}
\def\eea{\end{eqnarray}}

\def\BT{\begin{theorem}}
\def\ET{\end{theorem}}
\def\BL{\begin{lemma}}
\def\EL{\end{lemma}}
\def\BC{\begin{corollary}}
\def\EC{\end{corollary}}
\def\BE{\begin{example}}
\def\EE{\end{example}}
\def\BD{\begin{definition}}
\def\ED{\end{definition}}
\def\BR{\begin{remark}}
\def\ER{\end{remark}}
\def\BAS{\begin{assumption}}
\def\EAS{\end{assumption}}
\def\BAX{\begin{axiom}}
\def\EAX{\end{axiom}}
\def\BI{\begin{itemize}}
\def\EI{\end{itemize}}

\def\BMP{\begin{minipage}{9.5cm}}
\def\EMP{\end{minipage}}
\def\MPT{\begin{minipage}{11.5cm}}
\def\EPT{\end{minipage}}

\title{
{\normalsize CORE DISCUSSION PAPER }\\{\normalsize
2020/25}\\\vspace{10mm} \textbf{Online analysis of epidemics\\
with variable infection rate} }
\author{Yu. Nesterov
\thanks{Center for Operations Research and Econometrics (CORE),
Catholic University of Louvain (UCL). \newline E-mail:
Yurii.Neterov@uclouvain.be. ORCID 0000-0002-0542-8757. \newline
Research results presented in
this paper were obtained in the framework of ERC Advanced
Grant~788368.}}

\date{July 19, 2020
}

\begin{document}
\maketitle

\abstract{In this paper, we continue development of the new epidemiological model \cite{HIT}, which is suitable for analyzing and predicting the propagation of COVID-19 epidemics. This is a discrete-time model allowing a reconstruction of the dynamics of asymptomatic virus holders using the available daily statistics on the number of new cases. We suggest to use a new indicator, the total infection rate, to distinguish the propagation and recession modes of the epidemic. We check our indicator on the available data for eleven different countries and for the whole world. 
Our reconstructions are very precise.
In several cases, we are able to detect the exact dates of the disastrous political decisions, ensuring the second wave of the epidemics. It appears that for {\em all our examples} the decisions made on the basis of the current number of new cases are wrong. In this paper, we suggest a reasonable alternative. Our analysis shows that all tested countries are in a dangerous zone except Sweden.}


\thispagestyle{empty}

\newpage\setcounter{page}{1}

\section{Introduction}
\setcounter{equation}{0}

\vspace{1ex}\noindent
{\bf Motivation.} Devastating consequences of COVID-19 pandemic create new challenges not only for political institutions, but also for scientific communities. Important political decisions on the adequate response to the fast propagation of the virus must be based on scientific forecasts of the consequences. Wrong estimate of its actual dynamics can quickly transform a flourishing country into a highly polluted region. Six months of the history of this pandemic in the world provides us already with several examples of that type.

Unfortunately, this virus has some features, which are difficult to model by the classical epidemiological tools. Most of the epidemiological models are originated from the famous model SIR \cite{SIR}, proposed by Kermak and McKendrick in 1927. The most popular variants of this model are SEIR, SEIRS, and others \cite{SEIR0,Brauer,SEIR}.
We are not going to discuss now these models in details (see the monographs \cite{EBook3,EBook,EBook2}), since they have one common feature, making them inappropriate for COVID-19.

All elements of these models (susceptible, exposed, infected, recovered, and other individuals), are modeled as functions of {\em continuous time}, related by a system of ordinary differential equations (ODE). This means that the variations of all elements at time $t$ depends on the values of these elements at the {\em same moment} $t$. This is exactly what can hardly work with COVID-19. All observations of the last six months confirm that this virus has a significant contamination delay, from seven to fourteen days long. So, even a possibility of modeling its propagation by an ODE is questionable.

Another new situation with COVID-19 is a permanent flow of new data
describing the propagation of the virus over the world in the {\em real time}. We are living in the era of Big Data. And of course it is very interesting to use this data for enhancing our online prediction abilities. However, note that this is a {\em discrete-time} data, which is distributed in the form of daily reports. This means that it better fits the discrete-time dynamical systems, where it can be used directly without any preprocessing. 

These two observations served as the main motivation for developing a new discrete-time epidemiological model HIT \cite{HIT}, which fits well the main features of COVID-19. This is an axiomatic mathematical model, where its elements, number $H(d)$ of asymptomatic virus holders at day $d$, number of individuals $I(d)$ infected at day $d$, and the total number of cases $T(d)$, are functions of discrete time $d \in \Z$, related by some natural balance equations and one important assumption (axiom). Namely, we assume that the virus has a constant contamination delay $\Delta \geq 1$, where $\Delta$ is an integer parameter. This assumption allows us to prove the fundamental {\em Next-Day Law}:
$$
\ba{rcl}
H(d+1) & = & T(d+\Delta) - T(d), \quad d \in \Z,
\ea
$$
which relates the hidden characteristic $H(\cdot)$ with the observable function $T(\cdot)$. Consequently, we can relate the infection rate
$$
\ba{rcl}
\gamma(d) & \Def & {I(d) \over H(d)} \; = \; {T(d+\Delta) - T(d+\Delta-1) \over T(d+\Delta-1) - T(d-1)}, \quad d \in \Z,
\ea
$$
with function $T(\cdot)$ (see \cite{HIT}).

The main developments of \cite{HIT} were focused on the description of behavior of our characteristics during a strict containment period. During this time, it is natural to assume that the infection rate is piece-wise constant. This allows us to make short-term predictions for the future developments of function $T(\cdot)$. This idea was checked on the statistics of Belgium for the period March-May 2020. It was shown that the natural contamination delay for Belgium is $\Delta = 10$. Using this value of parameter, we showed that the online short-term prediction works very well, ensuring a very high accuracy in predicting the function $T(\cdot)$ and the hidden characteristic $H(\cdot)$. All of that needs a small number of updates for the constant level of the infection rate $\gamma(\cdot)$.

However, starting from June, many countries started to lift the lockdown restrictions. Consequently, the volatility of the infection rate was increased. Thus, it was necessary to develop a more stable aggregate characteristic, predicting the future behavior of our objects. In this paper, we suggest to look at the {\em total infection rate}
$$
\ba{rcl}
\Gamma(d) & \Def & \sum\limits_{i=0}^{\Delta-1} \gamma(d-i), \quad d \in \Z.
\ea
$$
It is possible to prove that if $\gamma(\cdot)$ is uniformly greater than one, than the characteristic $H(\cdot)$ grows with a linear rate. If it is uniformly smaller than one, then $H(\cdot)$ vanishes with the linear rate too. Hence, observing the behavior of the easily computable function $\Gamma(\cdot)$, we can detect the current mode of the development of our process.

This idea is verified in Section \ref{sc-App}, where we present an evolution
of this function for eleven different countries and for the whole world.
We show that this indicator reflects properly the actual evolution of the process. Moreover, we can even see the {\em exact dates} of events, resulted in the dramatic changes of the whole dynamics. Of course, all this information becomes available within a delay of $\Delta=10$ days. However, now it is clear that COVID-19 pandemic is a long-term process. So, this delay is not very essential.

\vspace{1ex}\noindent
{\bf Contents.} The paper is organized as follows, In Section \ref{sc-HIT}, we introduce the HIT model and prove that the total infection rate $\Gamma(\cdot)$ can be used for detecting the current status of the epidemic (propagation/recession). 

In Section \ref{sc-App}, we apply this indicator to the real-life data on COVID-19 developments in different countries. All our data is taken from the libraries \cite{DATA,Stat} up to the last date of July 18. Thus, we can reconstruct the developments of the epidemics up to July 8.

We consider the following countries.
\BI
\II
In Section \ref{sc-Bel}, we analyze the situation in Belgium. Our conclusion is that during the whole month of June, the total infection rate was fluctuating around the Epidemic Threshold $\gamma = 1$. And starting from the first days of July this indicator became growing. In our model, this is a clear sign of the second wave.
\II
In Section \ref{sc-Lux}, we consider the situation in Luxembourg. In accordance to our model, this country has already experienced the second wave of infection. At this moment, the total infection rate goes down, and very soon it must be below the threshold.
\II
Section \ref{sc-Germ} is devoted to Germany. This country demonstrates one of the most systematic way of fighting with COVID-19. However, it total infection rate is still in the neighborhood of the threshold, allowing some chances for the second wave.
\II
In Section \ref{sc-Ital}, we look at the situation in Italy. In accordance with our model, this country was very successful in fighting with the infection. However, after reopening the business activities, cinemas, and shops, the country is again in a dangerous zone. For avoiding the second wave, it can allow now no more than 200 new cases per day.
\II
In Section \ref{sc-Isr}, we look at the dynamics in Israel. Our model shows that this country was very successful in the beginning. However, after a victorious speech of its Prime Minister on May 7, all restrictions were lifted, and our main indicator started to grow very quickly. This was  realized too late. As a result, the country is fighting now with the second wave, which is much stronger than the first one.
\II
In Section \ref{sc-Jap}, devoted to Japan, we see the same story: inaccurate actions of the politicians resulted in the second wave.
\II
In the next Section \ref{sc-Braz}, we apply our model to Brazil, which is a big federal state. Therefore, we get a very regular picture showing the gradual decrease of the infection rate and stabilization of the total number of asymptomatic virus holders. We cannot see any signs of the recession yet.
\II
Is Section \ref{sc-US}, we consider the situation in the US. Again, since this is a big country, the whole picture is quite regular. We have seen a slow deterioration in the characteristic $H(\cdot)$ up to the end of May. However, starting from the second week of June, this indicator is increasing showing no intention to stabilize.
\II
Finally, in Section \ref{sc-World}, we apply our model to the statistics of the whole planet. The resulting picture is very regular, showing the monotone increase of the function $H(\cdot)$. At this moments, it is on the level of 2.1 millions. At the same time, the variations in the total infection rate correctly describe re-attribution of the leading role from China to Europe, and to USA with Latin America.
\II
In the last three sections we consider countries with nontrivial lockdownn strategies. In Section \ref{sc-Spain}, we look at Spain, the country which had a difficult but successful fighting against the first wave of infection. However, our model shows that it has now good chances to lose everything after opening the boarders for tourists.

In Section \ref{sc-Swed}, we look at Sweden, which did not introduce the strict lockdown rules. Nevertheless, our model shows that a responsible behavior of population is essential for a final success.

And in Section \ref{sc-Holl}, we consider The Netherlands, where the Intelligent Lockdown rules where applied. It was also a success story up to the moment of lifting on July~1st the travel restrictions. After this date, our model shows a permanent growth of the total infection rate, which already passed the Epidemic Threshold.
\EI

We conclude the paper with Section \ref{sc-Conc}, were we summarize our observations and discuss abilities of our model.

\section{HIT model with variable infection rate}\label{sc-HIT}
\SetEQ

This model is based on three dynamic characteristics of
epidemic development, which are functions of discrete time. Current situation at day $d \in \Z$
is described by the following objects.
\BI
\II
$H(d)$, the number of asymptomatic virus holders in the
beginning of day $d$. These holders infect other people
with certain rate. The rate and duration of the latent
period are also parameters of our model. They will be
introduced later.
\II
$I(d)$, the number of newly infected persons during the
day $d$. We call it Daily Infection Volume.
\II
$T(d)$, the total number of confirmed cases of infection
up to the end of the day $d$.
\EI
Among these characteristics, only
function $T(\cdot)$ is observable, at least in a
retrospective way. The two other functions will be
computed as an output of our model. In our considerations, it
is convenient to use also the following secondary
characteristic:
\BI
\II
$C(d)$, the number of new cases discovered at day $d$.
During the pandemic of COVID19, this information was
reported every day by the official sources.
\EI

Thus, we want to predict the behavior of these
discrete-time dynamic characteristics of propagation of
the disease, leaving the modeling of other important
characteristics (e.g. recovery and mortality rates) aside.
It is convenient to extend them onto all $d \in \Z$, having in mind
that for a deep past all of them are equal to zero. The
starting day of our model $d=0$ corresponds to the first
positive measurement of $T(\cdot)$.

The above characteristics satisfy two simple balance
equations, which follow from the physical meaning of our objects and the
standard way of functioning of the health system.
\BI
\II
{\em Conservation Law}
\beq\label{eq-CL}
\ba{rcl}
T(d) & = & T(d-1) + C(d), \quad d \in \Z.
\ea
\eeq
\II
{\em Propagation Law}
\beq\label{eq-PL}
\ba{rcl}
H(d+1) & = & H(d) + I(d) - C(d), \quad d \in \Z.
\ea
\eeq
This means that all newly detected asymptomatic virus
holders are isolated. The newly infected virus holders
become active only at the next day.
\EI

In HIT model, it is convenient to introduce one more
measurement.
\BD
Function $\gamma(d) \geq 0$, defined by the equation
\beq\label{eq-DIR}
\ba{rcl}
I(d) & = & \gamma(d) H(d), \quad d \in \Z,
\ea
\eeq
is called the {\em Daily Infection Rate} of our process.
If $H(d) = 0$, then $I(d)$ is also null, and we define
$\gamma(d)=0$.
\ED

Parameter $\gamma(d)$ is related to the
communication mode in the society at day $d$, which
influences the propagation rate of the virus. It
corresponds to the average number of persons infected
during one day by one asymptomatic virus holder. Hence,
it could be treated as expectation of a random variable.
However, in this paper we prefer to use a deterministic
terminology.

Finally, we need to introduce the last crucial assumption.
\BAX\label{ass-CLP}
Propagation of the virus has a \underline{\em Constant
Latent Period}. This means that
\beq\label{eq-CLP}
\ba{rcl}
C(d) & = & I(d-\Delta), \quad d \in \Z,
\ea
\eeq
where $\Delta \geq 1$ is an integer parameter. It is
called {\em Contamination Delay}.
\EAX

Since $C(d) = 0$ for $d \leq -1$, this axiom immediately
leads to the following consequences:
\beq\label{eq-INull}
\ba{rcl}
I(d) & = & 0, \quad d \leq - \Delta - 1,\\
\\
H(d) & \refEQ{eq-PL} & 0, \quad d \leq -\Delta.
\ea
\eeq

Now we have five discrete-time functions $H(\cdot)$, $I(\cdot)$, $T(\cdot)$, $C(\cdot)$, and $\gamma(\cdot)$, which are related by four equations (\ref{eq-CL}), (\ref{eq-PL}), (\ref{eq-DIR}), and (\ref{eq-CLP}). Hence, the knowledge of one of these functions allows the {\em exact} reconstruction of all others. Let us present three important links, having in mind that the function $T(\cdot)$ is observable.

{\bf 1. Next-Day Law} (see Theorem 1 in \cite{HIT}). For all $d \in \Z$, we have 
\beq\label{eq-NDLaw}
\ba{rcl}
H(d+1) & = & T(d+\Delta) - T(d).
\ea
\eeq

{\bf 2. Link {\boldmath$T(\cdot) \leftrightarrow \gamma(\cdot)$}} (see Theorem 2 in \cite{HIT}). For all $d \in \Z$, we have
\beq\label{eq-TGamma}
\ba{rcl}
T(d+1) & = & T(d) + \gamma(d-\Delta+1) \cdot (T(d) - T(d-\Delta)).
\ea
\eeq

{\bf 3. Link {\boldmath$H(\cdot) \leftrightarrow \gamma(\cdot)$}}. For all $d \in \Z$, we have
\beq\label{eq-HGamma}
\ba{rcl}
H(d+1) & = & \sum\limits_{k=d-\Delta+1}^d \gamma(k) H(k).
\ea
\eeq
(Indeed, by Theorem 1 in \cite{HIT}, $H(d+1) = \sum\limits_{k=d-\Delta+1}^d I(k)$, and (\ref{eq-HGamma}) follows from (\ref{eq-DIR}).)

In paper \cite{HIT}, it was shown that we can predict the future development of $T(\cdot)$ during the containment period, using the equation (\ref{eq-TGamma}). This is possible since the behavioral restrictions in the society result in a piece-wise constant shape of the infection rate $\gamma(\cdot)$. In this paper, we are more interested in pre- or post-containment periods, when the restrictions are not so severe. As a result, the volatility of this function becomes significant. It can depend, for example, on the level of professional activities or social communications typical for particular days of the week. Vacations, holidays, cultural events, and many other factors make behavior of the infection rate very irregular. Hence, for detecting the current mode of epidemics, we need to use some aggregated characteristics.

In this paper, we present some advancements in this direction. Our target characteristic is the unobservable function $H(\cdot)$. Its increasing behavior corresponds to the propagation mode of the epidemic. If this function decreases, then the epidemic is in a recession mode. Let us derive some sufficient conditions for identifying these two cases.

The main idea of our approach is based on the fact that the infection rate $\gamma(\cdot)$ can be recovered with certain delay by the observable function $T(\cdot)$:
\beq\label{eq-GammaT}
\ba{rcl}
\gamma(d) & \refEQ{eq-TGamma} & {T(d+\Delta) - T(d+\Delta-1) \over T(d+\Delta-1) - T(d-1)}, \quad d \in \Z.
\ea
\eeq
This means that we are able to compute the coefficients of the recurrence (\ref{eq-HGamma}), and use them for detecting the current mode of epidemic.

Let us define an aggregate characteristic, motivated by the representation (\ref{eq-HGamma}).
\BD\label{def-TIR}
Discrete-time function 
\beq\label{eq-TIR}
\ba{rcl}
\Gamma(d) & = & \sum\limits_{i=0}^{\Delta-1} \gamma(d-i)
\ea
\eeq
is called the {\em total infection rate} at day $d \in \Z$.
\ED
Clearly, the volatility of this function is much less than volatility of the daily infection rate $\gamma(\cdot)$. Moreover, looking at its behavior, we can make certain predictions on the future developments.

The next two theorems significantly enforce the results of Lemma 1 in \cite{HIT}.
\BT\label{th-Recc}
Let us assume that $\Gamma(d) \leq \sigma_r < 1$ for all $d \geq d_0$. Then
\beq\label{eq-Recc}
\ba{rcl}
H(d_0 + k) & \leq & C_r \cdot \sigma_r^{k/\Delta}, \quad k \geq 0,
\ea
\eeq
where $C_r =  \max\limits_{0 \leq i \leq \Delta-1} H(d_0-i) \sigma_r^{i/\Delta}$.
\ET
\proof
In view of our assumption, the optimal value $\tau_*$ of the following problem
$$
\ba{rcl}
\tau_* & = & \sup\limits_{\tau \geq 0} \Big\{ \tau: \; \sum\limits_{i=1}^{\Delta} \gamma(d-i+1) \tau^{i/\Delta} \leq 1, \; \forall d \geq d_0 \Big\}
\ea
$$
is at least one. Let us prove that it is strictly greater than one. Note that all functions $\tau^{i/\Delta}$, $i = 1, \dots, \Delta$, are concave and increasing in $\tau \geq 0$. Therefore,
$$
\ba{rcl}
\tau^* & \geq & \sup\limits_{\tau \geq 0} \Big\{ \tau: \; \sum\limits_{i=1}^{\Delta} \gamma(d-i+1) \left[ 1 + {i \over \Delta}(\tau - 1)\right] \leq 1, \; \forall d \geq d_0 \Big\}\\
\\
& \geq & \sup\limits_{\tau \geq 0} \Big\{ \tau: \; \sum\limits_{i=1}^{\Delta} \gamma(d-i+1) \cdot \tau \leq 1, \; \forall d \geq d_0 \Big\}\;
\geq \; \sigma_r^{-1}.
\ea
$$
Denoting now $r = \sigma_r^{1/\Delta}<1$, we have
$\sum\limits_{i=1}^{\Delta} \gamma(d-i+1) \cdot r^{-i} \leq 1$ for all $d \geq d_0$. In other words,
\beq\label{eq-R}
\ba{rcl}
\sum\limits_{i=0}^{\Delta-1} \gamma(d-i) \cdot r^{-i} & \leq & r, \quad \forall d \geq d_0.
\ea
\eeq

Assume now that for some $d \geq d_0$ 
we have
\beq\label{eq-Ind1}
\ba{rcl}
H(d-i) & \leq & C r^{d-i}, \quad i = 0, \dots \Delta-1.
\ea
\eeq
with some $C > 0$. Then 
$$
\ba{rcl}
H(d+1) & \refLE{eq-HGamma} & \sum\limits_{i=0}^{\Delta-1} \gamma(d-i) \cdot C r^{d-i} \; \refLE{eq-R} C r^{d+1}.
\ea
$$
Thus, we prove inequality (\ref{eq-Ind1}) for all $d \geq d_0$ provided that we have chosen $C$ as follows:
$$
\ba{rcl}
H(d_0-i) & \leq & C r^{d_0-i}, \quad i = 0, \dots \Delta-1.
\ea
$$
This means that $C = \max\limits_{0 \leq i \leq \Delta-1} H(d_0-i)r^{i - d_0} \equiv r^{-d_0} C_r$.
\qed

Let us present now a sufficient condition for propagation mode of the epidemic.
\BT\label{th-Prop}
Let us assume that $\Gamma(d) \geq \sigma_p > 1$ for all $d \geq d_0$. Then
\beq\label{eq-Prop}
\ba{rcl}
H(d_0 + k) & \geq & C_p \cdot r_p^{k}, \quad k \geq 0,
\ea
\eeq
where $r_p = 1 + {1 \over \Delta} \left(1 - \sigma_p^{-1}\right) > 1$, and $C_p =  \min\limits_{0 \leq i \leq \Delta-1} H(d_0-i) r_p^i$.
\ET
\proof
In view of our assumption, the optimal value $r_*$ of the following problem
$$
\ba{rcl}
r_* & = & \sup\limits_{r > 0} \Big\{ r: \; \sum\limits_{i=0}^{\Delta-1} \gamma(d-i) r^{-i-1} \geq 1, \; \forall d \geq d_0 \Big\}
\ea
$$
is at least one. Let us prove that it is strictly greater than one. Note that all functions $r^{-i-1}$, $i = 0, \dots, \Delta-1$, are convex and decreasing in $r > 0$. Therefore,
$$
\ba{rcl}
r_* & \geq  & \sup\limits_{r > 0} \Big\{ r: \; \sum\limits_{i=0}^{\Delta-1} \gamma(d-i) \left[1 - (i+1)(r-1) \right] \geq 1, \; \forall d \geq d_0 \Big\}\\
\\
& \geq & \sup\limits_{r > 0} \Big\{ r: \; \sum\limits_{i=0}^{\Delta-1} \gamma(d-i) \left[1 - \Delta(r-1) \right] \geq 1, \; \forall d \geq d_0 \Big\}\\
\\
& \geq & \sup\limits_{r > 0} \Big\{ r: \; \sigma_p \left[1 - \Delta(r-1) \right] \geq 1, \; \forall d \geq d_0 \Big\} \; = \; r_p.
\ea
$$
Thus, we have
\beq\label{eq-RS}
\ba{rcl}
\sum\limits_{i=0}^{\Delta-1} \gamma(d-i) \cdot r_p^{-i} & \geq & r_p, \quad \forall d \geq d_0.
\ea
\eeq

Assume now that for some $d \geq d_0$ 
we have
\beq\label{eq-Ind2}
\ba{rcl}
H(d-i) & \geq & C r_p^{d-i}, \quad i = 0, \dots \Delta-1.
\ea
\eeq
with some $C > 0$. Then 
$$
\ba{rcl}
H(d+1) & \refGE{eq-HGamma} & \sum\limits_{i=0}^{\Delta-1} \gamma(d-i) \cdot C r_p^{d-i} \; \refGE{eq-RS} C r_p^{d+1}.
\ea
$$
Thus, we prove inequality (\ref{eq-Ind2}) for all $d \geq d_0$ provided that we have chosen $C$ as follows:
$$
\ba{rcl}
H(d_0-i) & \geq & C r_p^{d_0-i}, \quad i = 0, \dots \Delta-1.
\ea
$$
Hence, we can take $C = \min\limits_{0 \leq i \leq \Delta-1} H(d_0-i)r_p^{i - d_0} \equiv r_p^{-d_0} C_p$.
\qed

Note that characteristic $\Gamma(d)$ can be computed by the expressions (\ref{eq-TIR}), (\ref{eq-GammaT}) only at the day $d+\Delta$, when all values $T(k)$ with $k \leq d+\Delta$ become known. Nevertheless, its values and its dynamics can help us in understanding the future trends of the epidemic. In the next section, we present some observations on the developments of the pandemic COVID-19 in 2020 in different countries.

\section{Application examples}\label{sc-App}
\SetEQ

Let us compare the behavior of the dynamic estimate $\Gamma(\cdot)$ with actual development of COVID-19 epidemics in different countries. The last day of our observations is July~18, 2020. In order to use the formulas (\ref{eq-GammaT}), (\ref{eq-TIR}), we need to choose the length of contamination delay. In \cite{HIT}, it was shown that its correct value for Belgium is $\Delta = 10$. Of course, for each country this delay can be different. However, at this moment we do not have enough information and human resources for investigating this question in details. Therefore, for the first attempt of application of our model in the world, we fix $\Delta = 10$.

\subsection{Belgium}\label{sc-Bel}

Let us continue our observations on development of the epidemic in Belgium. Its first three months, March, April, and May, 2020, were analyzed from the viewpoint of HIT model in \cite{HIT}. It was also shown that during the strict containment, the actual daily infection rate can replaced by a piece-wise linear function. This gives us a possibility for a short-term online forecast for the growth of function $T(\cdot)$ for the nearest future. In \cite{HIT} it was shown that the accuracy of this local prediction can be very high, typically better than $0.5\%$ in the relative scale.

It was also mentioned that in the end of May the estimated values of the infection rate $\gamma(\cdot)$ were approaching the dangerous threshold $\gamma = 1/\Delta = 0.1$. However, at that moment (beginning of June 2020), we did not have any justification for the total infection rate $\Gamma(\cdot)$. Now we can look at the whole history of the epidemic in Belgium for five months using this characteristic.

\begin{figure}[h]
\centering
\includegraphics[scale=0.5]{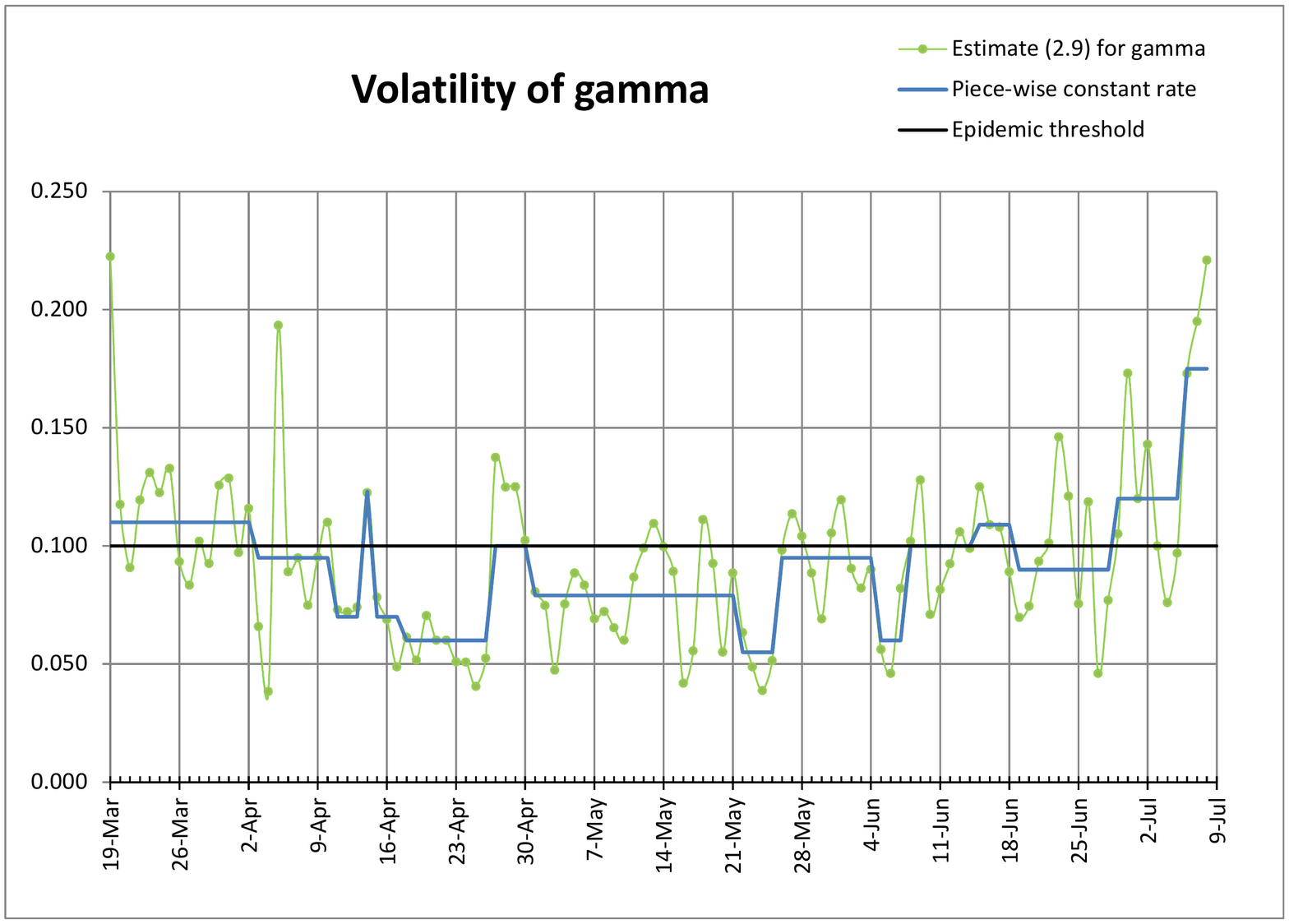} 
\caption{Comparison of daily estimates for $\gamma(\cdot)$ by (\ref{eq-GammaT}) with piece-wise constant function providing the same growth of $T(\cdot)$. Belgium, March-July 2020.}\label{fig-VolGamma}
\end{figure}

At the beginning of June 2020, Belgian government started the first de-containment measures, which clearly increased volatility of the daily infection rate. On Figure \ref{fig-VolGamma}, we can see that it becomes more and more difficult to trust the value $\gamma(d)$ for a particular day $d$. 

\begin{figure}[h!]
\centering
\includegraphics[scale=0.5]{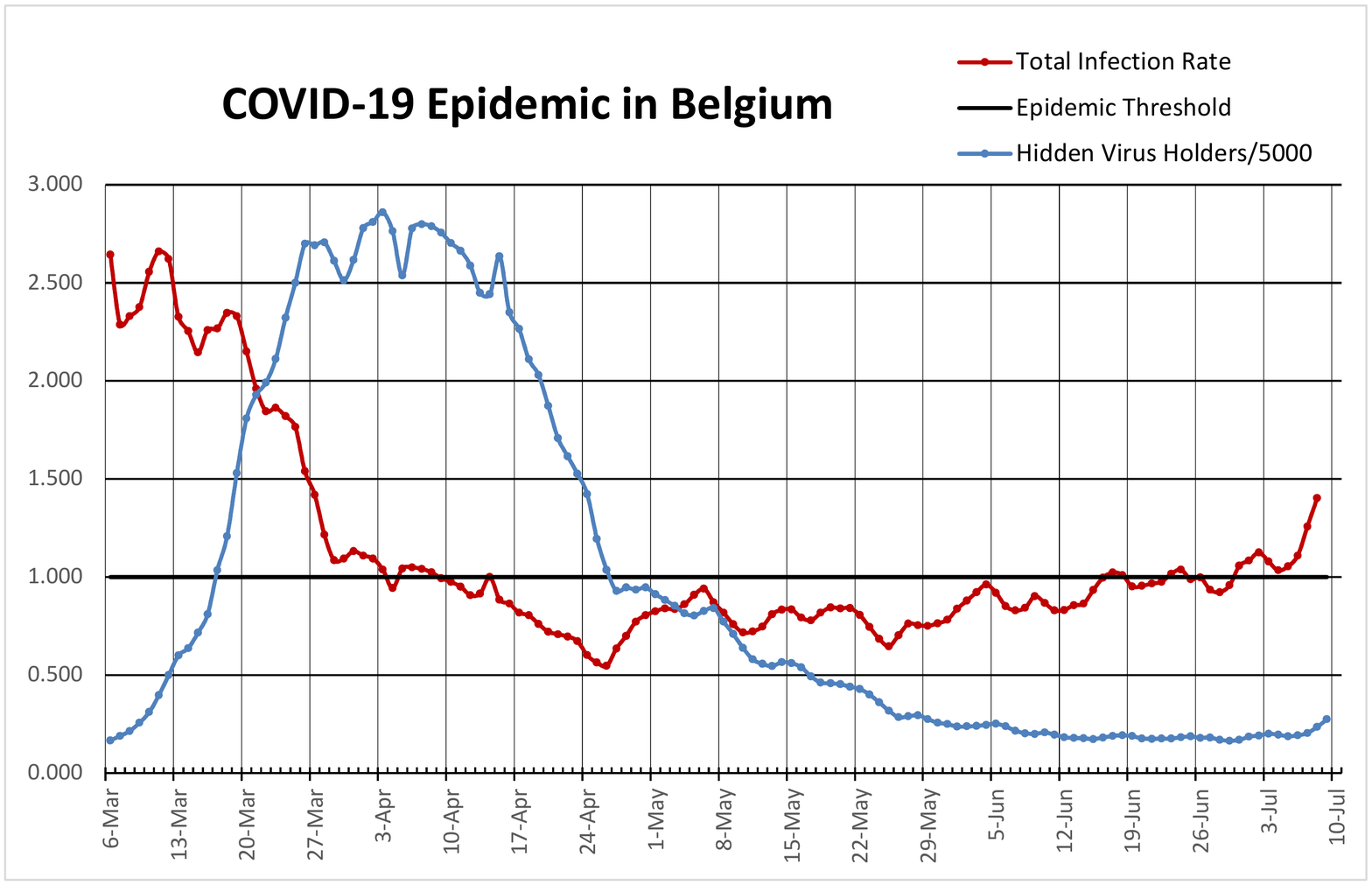} 
\caption{Development of the total infection rate $\Gamma(\cdot)$ in comparison with the estimate $H(\cdot)$ for the asymptomatic virus holders. Belgium, March-July 2020.}\label{fig-Growth}
\end{figure}

This is quite natural for a non-restricted behavior of people. Thus, in order to analyze the epidemic development in June and later, we need to look at the aggregate characteristic, the total infection rate $\Gamma(\cdot)$. Let us recall the main events in the epidemic history of Belgium and compare them with the evolution of our characteristics $\Gamma(\cdot)$ and $H(\cdot)$ displayed at Figure \ref{fig-Growth}.

The first case of COVID-19 was detected in Belgium in March 2, 2020. This infection was completely new for the country. Therefore, in the first week of March, the total infection rate was on the maximal level, typical for non-restricted behavior of population. Starting from March 13, we can see a small drop in the rate, which can be explained by better understanding of danger of this disease. 

The real drop in the values of $\Gamma(\cdot)$ started from March 19, the first full day of the containment. Its first two weeks were extremely successful, reducing the total infection rate almost up to the Epidemic Threshold $\gamma = 1$. After that, for the first two weeks of April, we see a stagnation of the rate around this value. It can be explained by a big number of asymptomatic virus holders, which was fluctuating around its maximal level 13500. 

The last two weeks of April were very good, ensuring a significant drop both in the function $H(\cdot)$ and in the total infection rate. The reasons for its termination are not clear yet. On one hand, it could be related on the massive return of Belgian citizens delayed by quarantine measures abroad. It was also the time for the first de-containment steps. Anyway, Belgium entered June by much worse values of $\Gamma(\cdot)$ (this was mentioned in Appendix of \cite{HIT}). During the whole month the total infection rate was fluctuating around the dangerous value. And in the beginning of July we see its strong growth. In accordance to our analysis (see Theorem \ref{th-Prop}), this should be a beginning of the second wave of infection. However, we can check this prediction only in the nearest future. At this moment, we have approximately the same number of asymptomatic virus holders as on March 6 (around 1300), but the current total infection rate is twice smaller. 

\subsection{Luxembourg}\label{sc-Lux}

This small country is very close to Belgium by their traditions and style of life. Therefore, we can expect that its parameter $\Delta$ is also close to ten. Let us look at the development of epidemics in this country.

As in all other countries, the beginning of epidemics was very aggressive. However, the reaction of authorities was also very fast. The schools were closed on March 16, and the containment has started. It was very efficient. In ten days the total infection rate was decreased below the Epidemic Threshold, and after April 6 the country kept the infection rate on a reasonably small level, which resulted in a very small number of asymptomatic virus holders. At May 4, this number is estimated as 82. 

\begin{figure}[h!]
\centering
\includegraphics[scale=0.5]{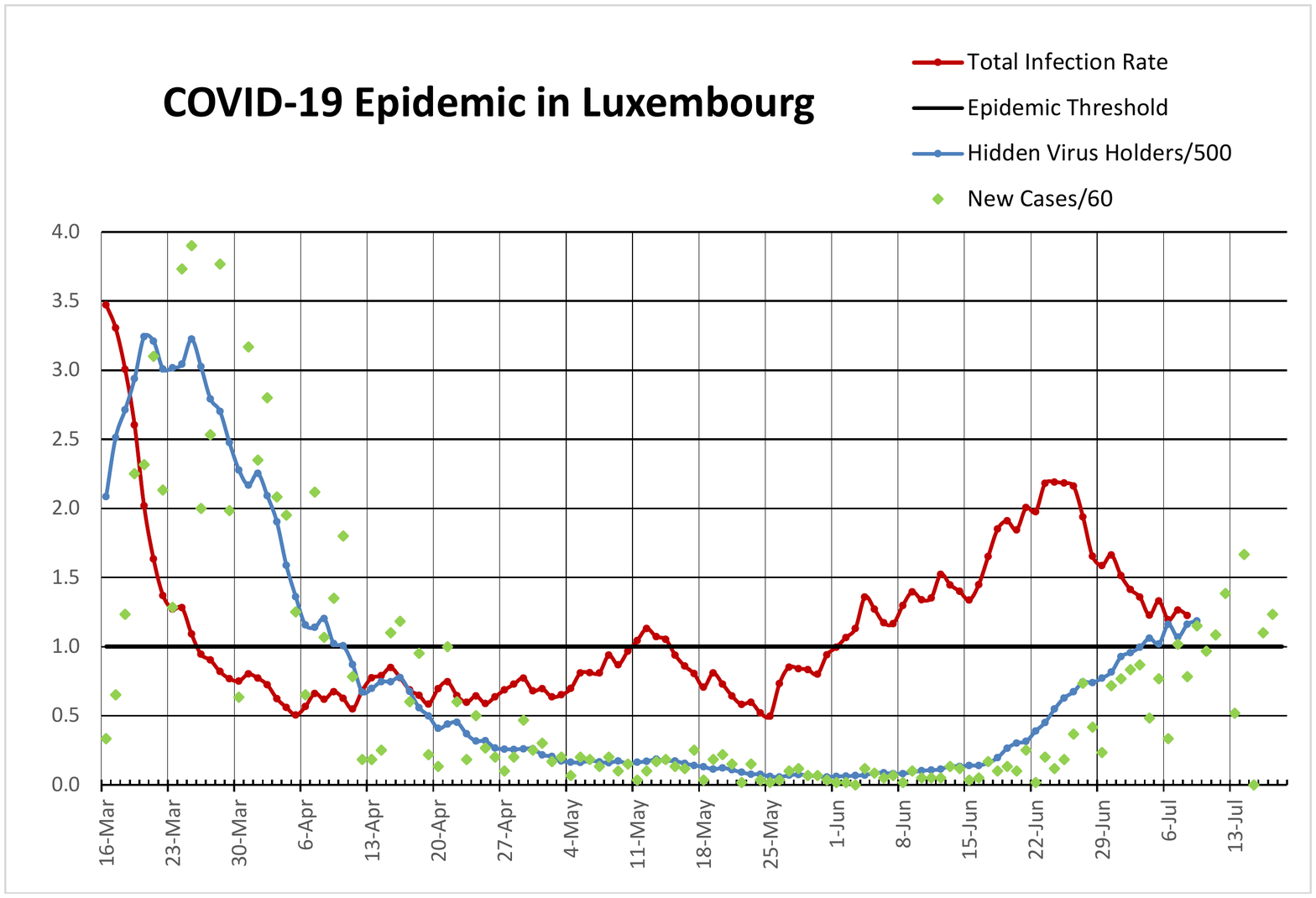} 
\caption{Development of the total infection rate $\Gamma(\cdot)$ in comparison with the estimate $H(\cdot)$ for the asymptomatic virus holders. Luxembourg, March-July 2020.}\label{fig-LuxGrow}
\end{figure}

At these days, the number of new cases was fluctuating around 10. Therefore, the de-containment measures of May 6 and May 11 looked very natural. However, we can easily find these point at Figure \ref{fig-LuxGrow}. May 4 marks the beginning of 10-days growth of the indicator $\Gamma(\cdot)$, and May 11 is a beginning of four-days period when the total infection rate was above the Epidemic Threshold. After that it became decreasing for the next two weeks. It is interesting that during all these days, the number of asymptomatic virus holders was also decreasing. It went down up to negligible 28 persons (on May 31st; recall that the population of Luxembourg is 626000). 

However, it was already at the moment when our indicator $\Gamma(\cdot)$ has crossed the threshold. Nothing was alarming at that time. The total number of new cases for the whole week June 1-7 was only 19. But the total infection rate definitely indicated the beginning of the second wave. And it indeed became visible very soon. 

In order to understand what happened on May 25, we need to live in this country. In any case, in four weeks, the growth of the number of new cases became so significant, that the authorities introduced again some re-containment measures (June 24). This date is also very well visible at Figure \ref{fig-LuxGrow}, as the maximal value of the total infection rate.

The further development of our indicator just confirms a high level of responsibility in this country. In the last days, we see a fast decrease of the infection rate, and there is no doubt that it will go below the Epidemic Threshold very soon. Note that our optimistic forecast is based on seemingly bad statistic for the number of new cases around the date July 13. However, this statistic corresponds to a {\em decreasing} behavior of the indicator $\Gamma(\cdot)$ around the date July 3.

This example shows that our approach is working even in the small countries, where the number of daily new cases is measured by dozens, not by thousands. Recall that all events shown at Figure \ref{fig-LuxGrow} are detectable by our indicator $\Gamma(\cdot)$ within a delay of ten days.

\subsection{Germany}\label{sc-Germ}

\begin{figure}[h!]
\centering
\includegraphics[scale=0.5]{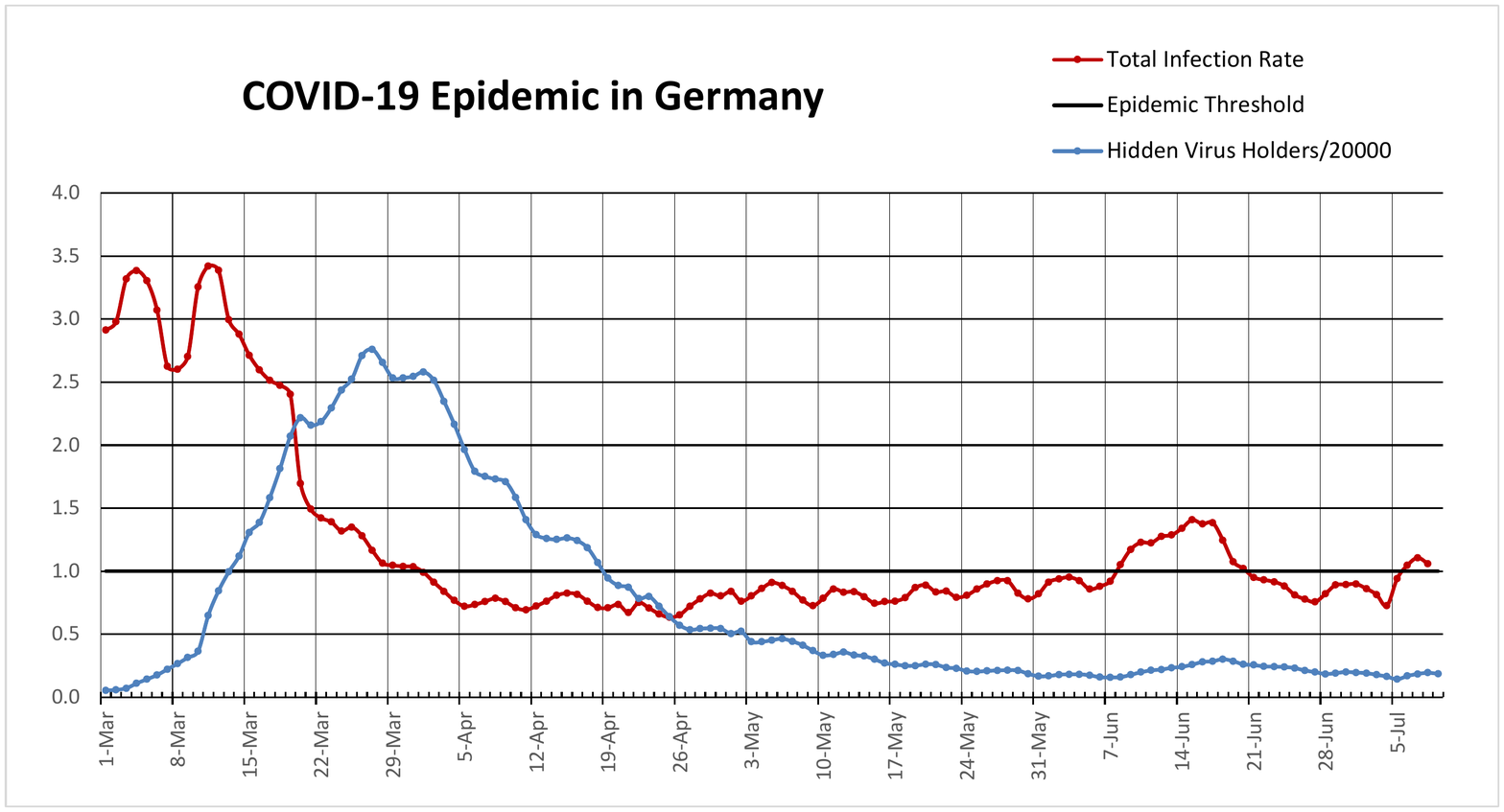} 
\caption{Development of the total infection rate $\Gamma(\cdot)$ in comparison with the estimate $H(\cdot)$ for the asymptomatic virus holders. Germany, March-July 2020.}\label{fig-GermGrow}
\end{figure}

Germany demonstrates one of the most systematic ways of treating the COVID-19 epidemic. The lockdown measures, gradually introduced in the period March 13-22, resulted in a quick drop of the total infection rate below the threshold.\footnote{The official lockdown was introduced in Germany on March 22. However, the first measures, like closing the schools, were activated from March 13. Since Germany is a federal state, there were also some local measures. Therefore, it is difficult to use here the advise \cite{HIT} of using the lockdown date for estimating the contamination delay $\Delta$. Hence, we keep $\Delta = 10$ for Germany too.}
After that, at Figure~\ref{fig-GermGrow}, we can see a regular behavior of the indicator $\Gamma(\cdot)$, with very slight deterioration. It is interesting that we observe also weekly cycles, which can be explained only by some economical activity, specific for different days of the week. From our model, we do not see any confirmation for several waves of fear on the beginning of the second wave, which were discussed in media in May. Only in the period of two weeks June 8-21, the total infection rate was above the threshold. However, this danger was detected (with certain delay) and eliminated by renewing the containment measures. Note that from our model this danger was visible starting from May 18, five days earlier as compared to the other methods.

Despite to its regularity, the situation in Germany remains uncertain since the total infection rate is very close to the threshold. It was never reduced up to the level of 0.5, which gives a hope for a fast elimination of the virus. At this moment, our estimate for the total number of asymptomatic virus holders in Germany is on the level 4000. With the current infection rate, its complete elimination needs many months. At the same time, 400 new cases per day will keep it constant.

\newpage

\subsection{Italy}\label{sc-Ital}

This country was the first victim of COVID-19 in Europe.
In accordance to our model, the fast growth of the total infection rate in February was stopped by the local containment measures and the national lockdown of March 9. The maximal number of hidden virus holders was close to 55000.
Next three weeks were very successful in reducing the infection rate. And then, during two months, the characteristic $H(\cdot)$ was gradually decreased. 

Starting from May 18, there were reopened most of the business activity, and on May~25, the swimming pools and gyms were also allowed. Theaters and cinemas are working since June 15. As a result, we can see the growth of the total infection rate, but fortunately only up to the threshold. Consequently, the total number of asymptomatic virus holders remains constant on the level of 2000. Thus, the safe number of new cases for Italy now is of the order 200 per day.

\begin{figure}[h!]
\centering
\includegraphics[scale=0.5]{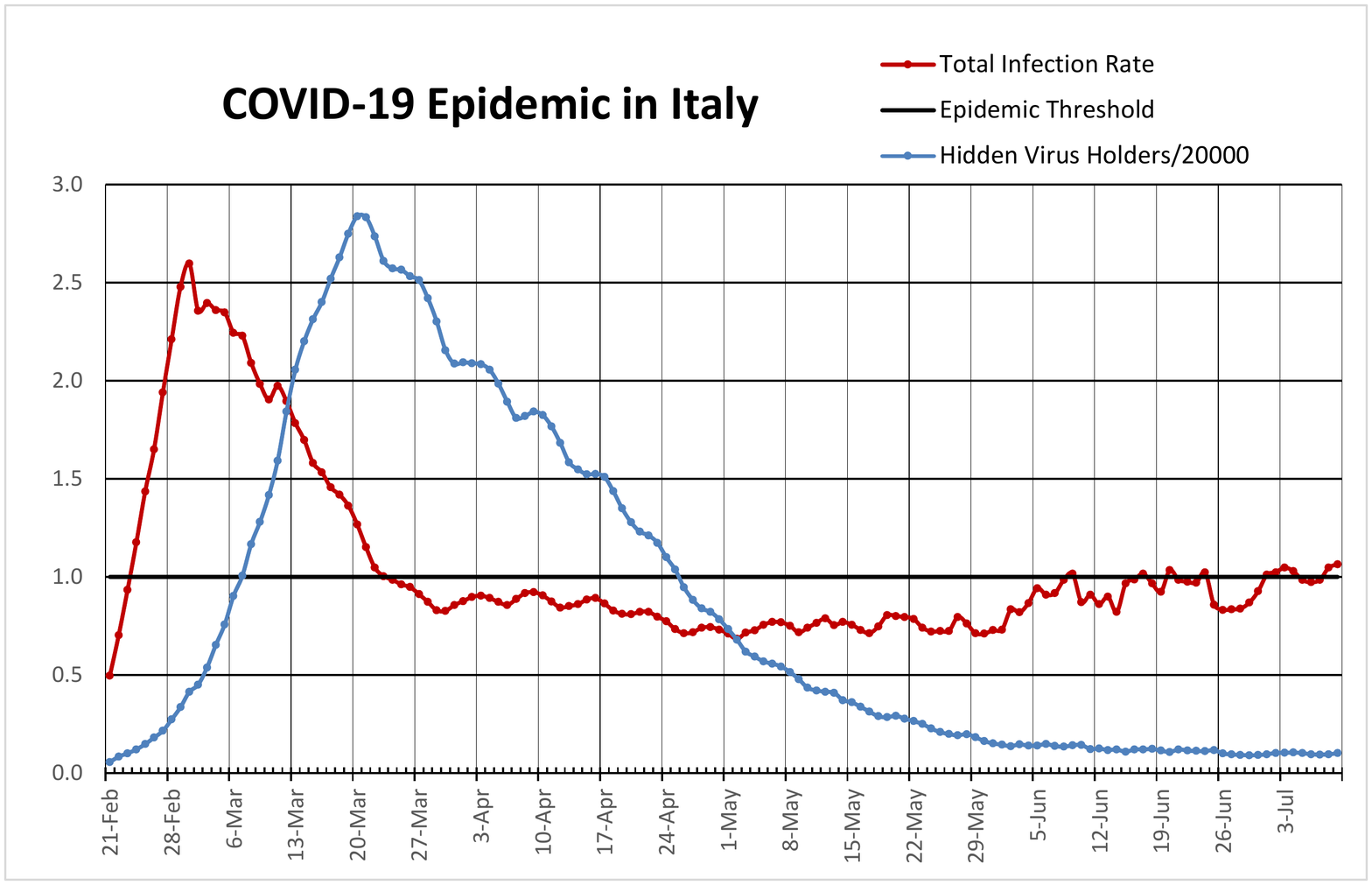} 
\caption{Development of the total infection rate $\Gamma(\cdot)$ in comparison with the estimate $H(\cdot)$ for the asymptomatic virus holders. Italy, February-July 2020.}\label{fig-ItalGrow}
\end{figure}

\subsection{Israel}\label{sc-Isr}

This country experiences now a big second wave of infection. In the beginning of the epidemic, actions of the government were very rapid. On March 11, Israel enforced social distancing and restricted the size of the meetings up to 100, and up to 10 from March~15. Moreover, on March 19, the Prime Minister declared a national state of emergency. These measures are properly recognized by our model. In three weeks, the total infection rate was decreased below the Epidemic Threshold, and number of hidden virus holders started to decrease from its local maximum of 5800 to a small number 141 on May 16. 

\begin{figure}[h!]
\centering
\includegraphics[scale=0.5]{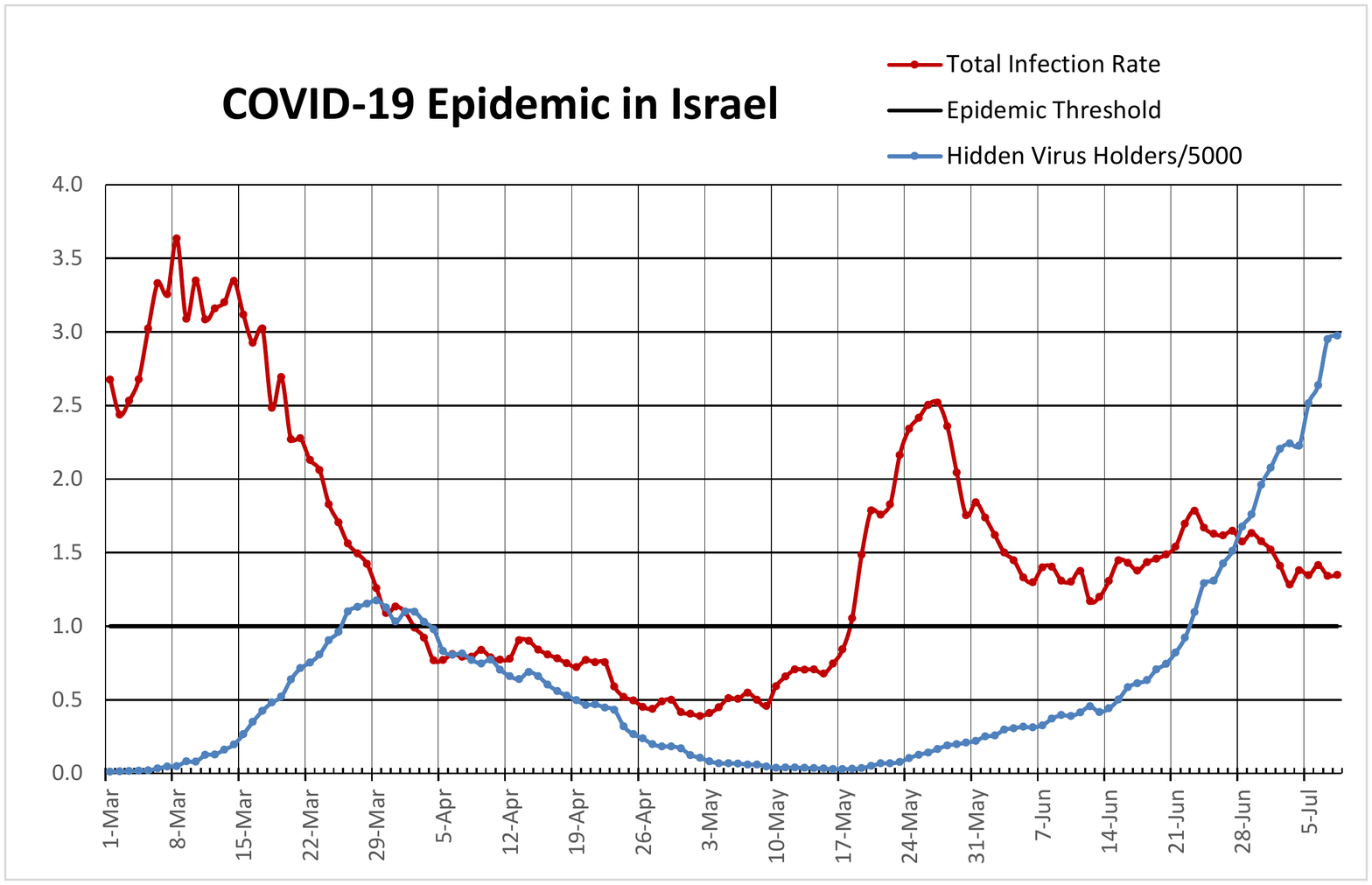} 
\caption{Development of the total infection rate $\Gamma(\cdot)$ in comparison with the estimate $H(\cdot)$ for the asymptomatic virus holders. Israel, March-July 2020.}\label{fig-IsrGrow}
\end{figure}

However, if we look at the total infection rate (see Figure \ref{fig-IsrGrow}), we can easily find a breaking point at May~9. After this date, it became constantly increasing for a long time, justifying the second wave of infection. From this observation, we could guess that something should happen around this date, something which changed the whole dynamics so dramatically. And indeed, on May 7 there was a national address of Prime Minister, who announced a further relaxation of coronavirus restrictions in Israel. At that moment, this decision was looking quite natural. Indeed, in this period the number of new cases per day was very low. For example, for two weeks of May 10-23, it was varying between 10 and 29. At the same time, in our model these two weeks correspond to a dramatic growth of the infection rate.  A significant growth in the daily statistics became visible only during the next two weeks. New restrictions dropped the rate down, but not sufficiently. Thus, now the total infection rate remains above the threshold, which results in the increase of the hidden virus holders. At this moment, we estimate this number close to 15000.

Note that with our model it was possible to detect the dangerous growth in the infection rate not later than in the end of May, when the total number of asymptomatic virus holders was less than one thousand.

\newpage 

\subsection{Japan}\label{sc-Jap}

This is the second country in our survey, fighting with the second wave right now. However, very recently, up to the end of June, Japan was considered as a very successful country. It had a small number of restrictions introduced by the government (see \cite{Time}).
There was no restrictions on residents movements and small businesses. No high-tech tracking systems were deployed. And even the number of tests was very low. Nevertheless, our model confirms that in the end of May the number of hidden virus holders was less than 400, which looks negligible for the country with population of 126 millions. 

The experts still do not understand how all of that was possible \cite{Time}. It seems that the main factor was the highest level of social responsibility of population combined with a strict following the simplest behavioral restrictions like social distancing, washing hands, etc. This is confirmed by the fact that at some moment the miracle was gone. 

\begin{figure}[h!]
\centering
\includegraphics[scale=0.45]{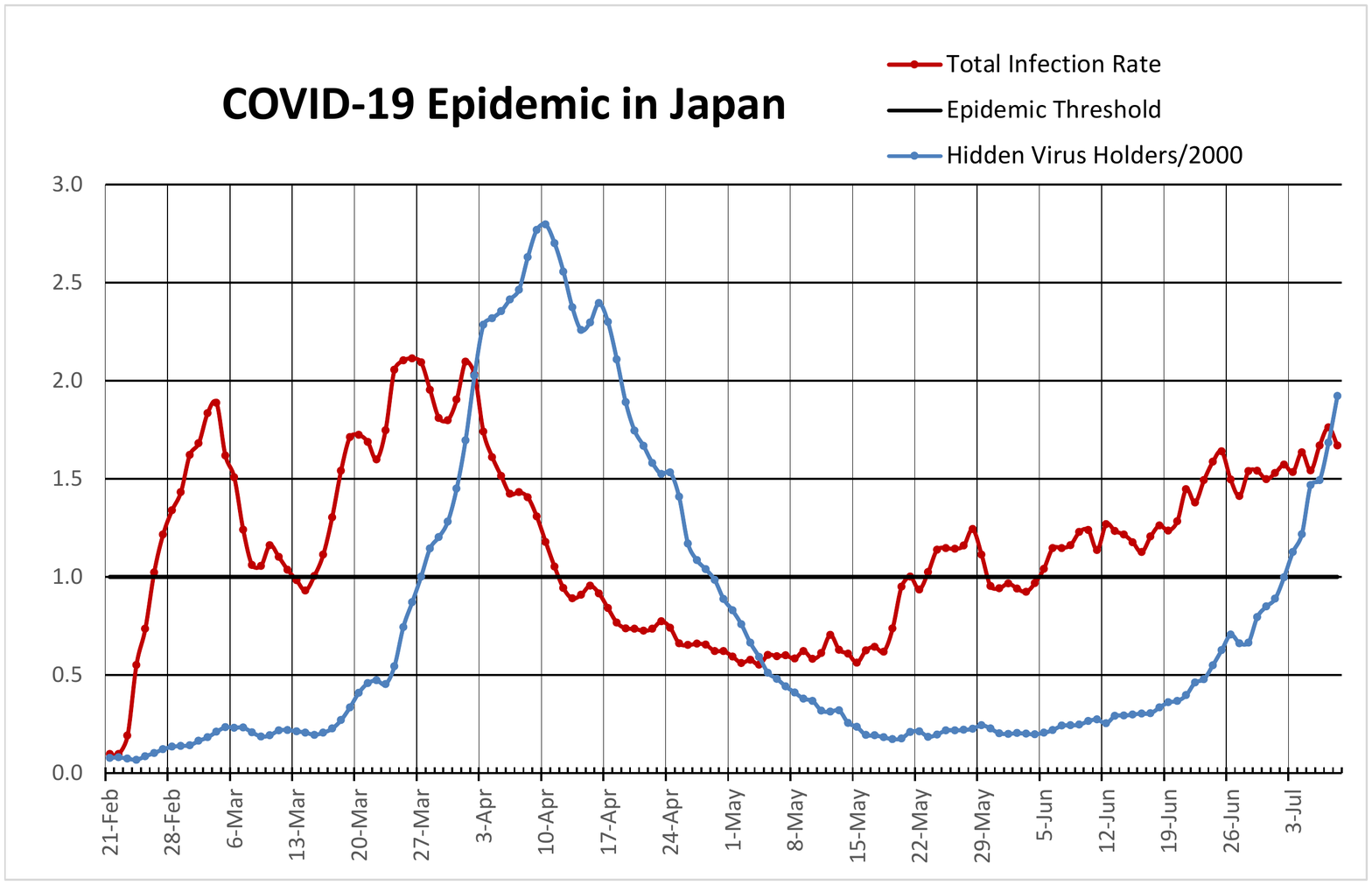} 
\caption{Development of the total infection rate $\Gamma(\cdot)$ in comparison with the estimate $H(\cdot)$ for the asymptomatic virus holders. Japan, February-July 2020.}\label{fig-JapGrow}
\end{figure}

Our model computes this breaking point as May 15 (see Figure \ref{fig-JapGrow}). This perfectly corresponds to the history: on May 14, the Prime Minister announced a victory in the battle with COVID-19 \cite{BBC}.
Starting from this date, Japan has lifted the state of emergency. After that, during one month nothing was alarming. Daily number of cases was fluctuating in the interval from 32 to 89. However, our model shows an {\em immediate growth} of the total infection rate. It was possible to see its passage above the threshold no later than on June 1st. At this moment, the number of asymptomatic virus holders was still of the order of four hundreds. Clearly, the restoration of social restrictions at that moment could be much less painful than now, when the level of $H(\cdot)$ is of  the order of four thousands.

\newpage

\subsection{Brasil}\label{sc-Braz}

In our survey, we come to the level of big federal states. Of course, Germany is a federal state too. However, the level and style of life in its different parts is very homogeneous. Brazil is twice bigger than Germany, and the diversity in the life there is much bigger. Let us look how our model can handle statistics of such a big country.

\begin{figure}[h!]
\centering
\includegraphics[scale=0.5]{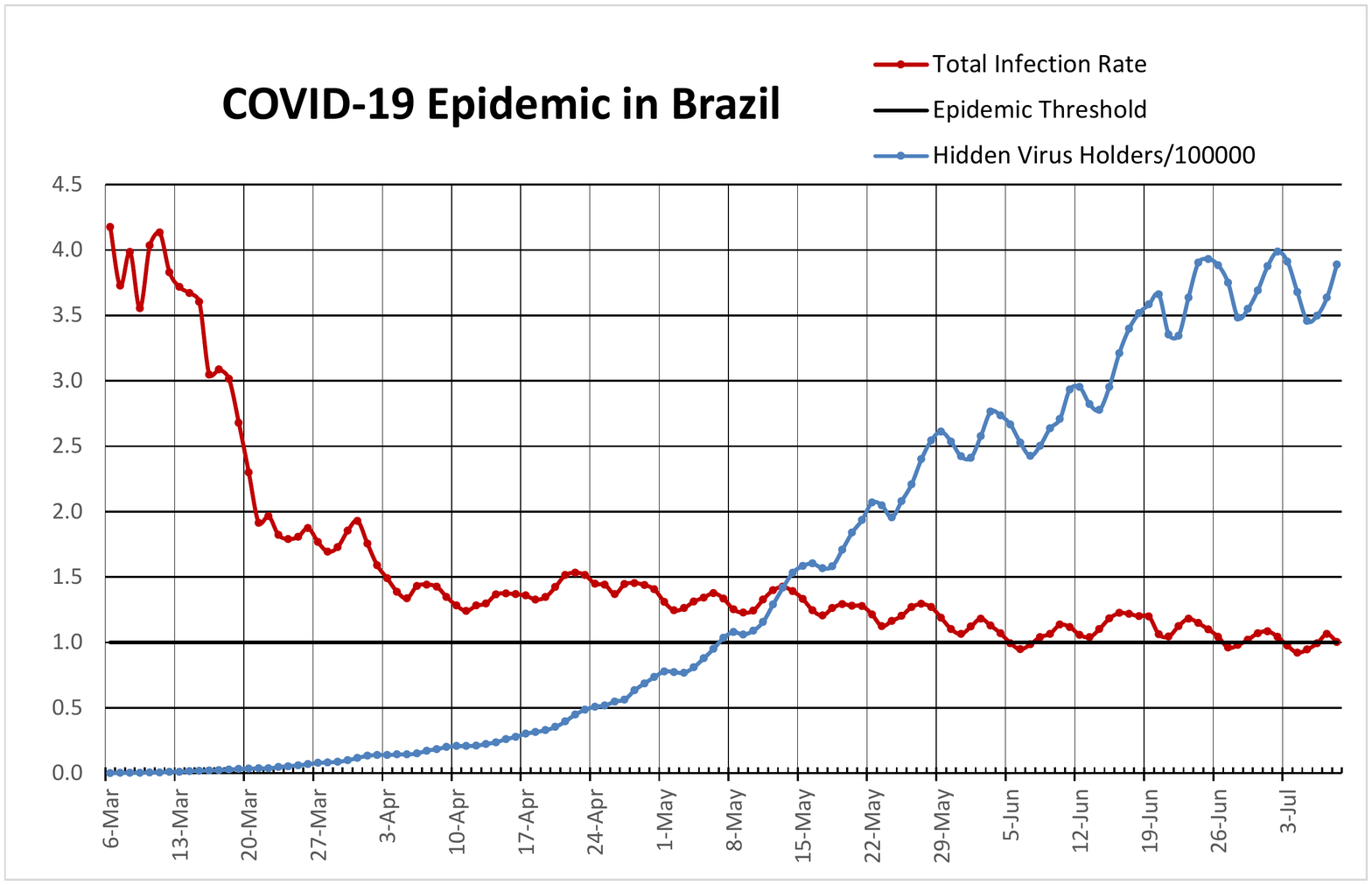} 
\caption{Development of the total infection rate $\Gamma(\cdot)$ in comparison with the estimate $H(\cdot)$ for the asymptomatic virus holders. Brazil, March-July 2020.}\label{fig-BraGrow}
\end{figure}

Surprisingly enough, our characteristics for Brazil demonstrate quite a regular behavior.
At Figure \ref{fig-BraGrow}, we can see that our model correctly computes the switching period of March~17-20, when the main municipalities of Brazil declared the state of emergency. After that, the total infection rate is gradually decreasing, approaching the Epidemic Threshold from above. Hence, our estimate for the number of asymptomatic virus holders in the country is fluctuating in the interval 350-400 thousands. So, it seems that the epidemic is entering a stagnation mode. In order to cross the threshold, more restrictions in social life are needed. However, this is not easy because of the current political situation.

\subsection{The United States}\label{sc-US}

This is one more federal state, with very high diversity in the style of life and difficult political situation. Our model correctly detects the period of activating "stay at home" quarantines in the main states. 
Starting from April 5, the total infection rate was fluctuating around the Epidemic Threshold, reducing at the same time the total number of hidden virus holders. However, starting from June 7, the infection rate became growing, in parallel with the massive political protests in the US. Consequently, the total number of hidden virus holders is growing very rapidly till now, exceeding the level of 600 thousands. Most probably, it can be reduced only by some lockdown measures. 

\begin{figure}[h!]
\centering
\includegraphics[scale=0.5]{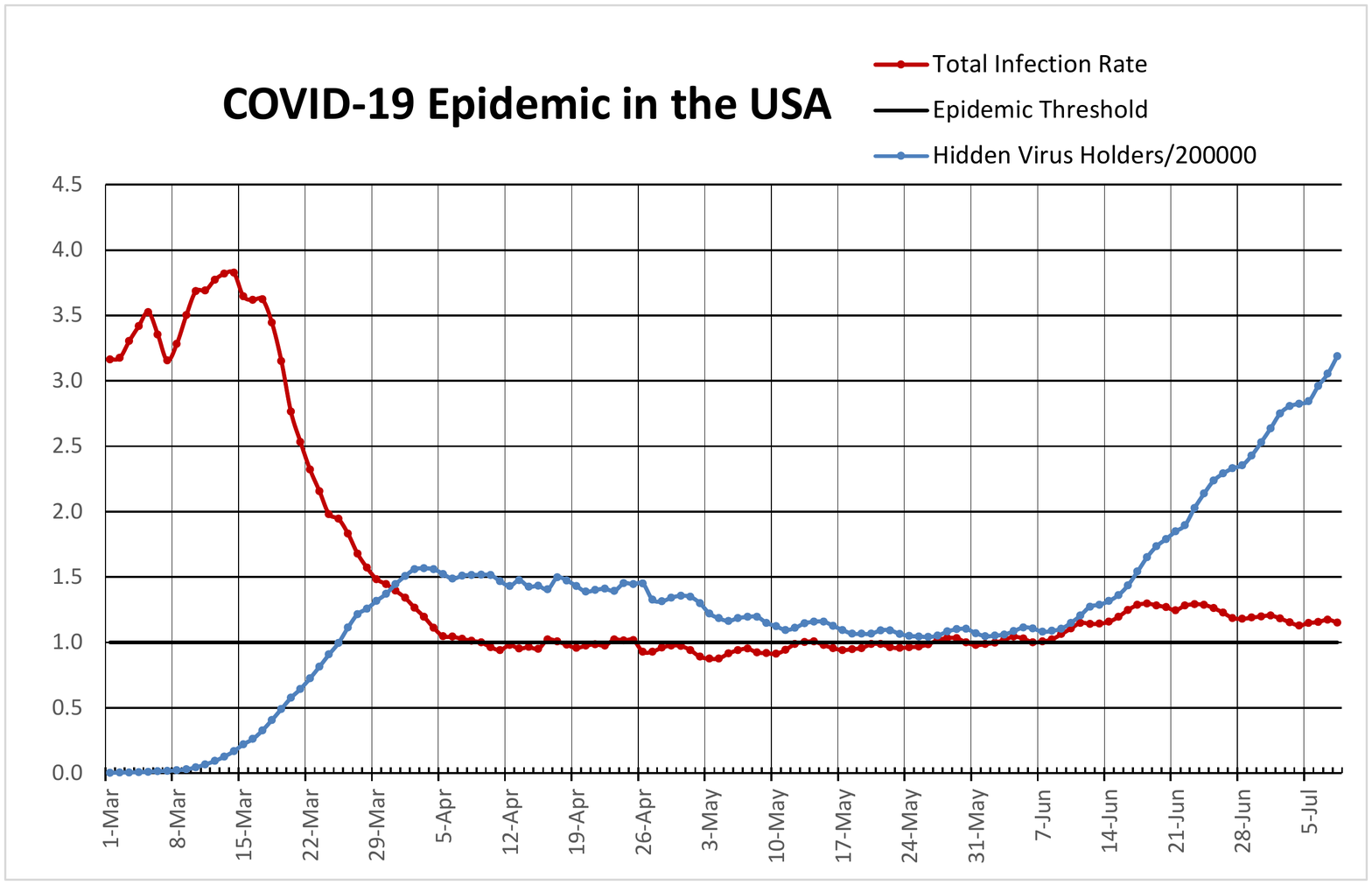} 
\caption{Development of the total infection rate $\Gamma(\cdot)$ in comparison with the estimate $H(\cdot)$ for the asymptomatic virus holders. The USA, March-July 2020.}\label{fig-USAGrow}
\end{figure}

\subsection{The World}\label{sc-World}

\begin{figure}[h!]
\centering
\includegraphics[scale=0.5]{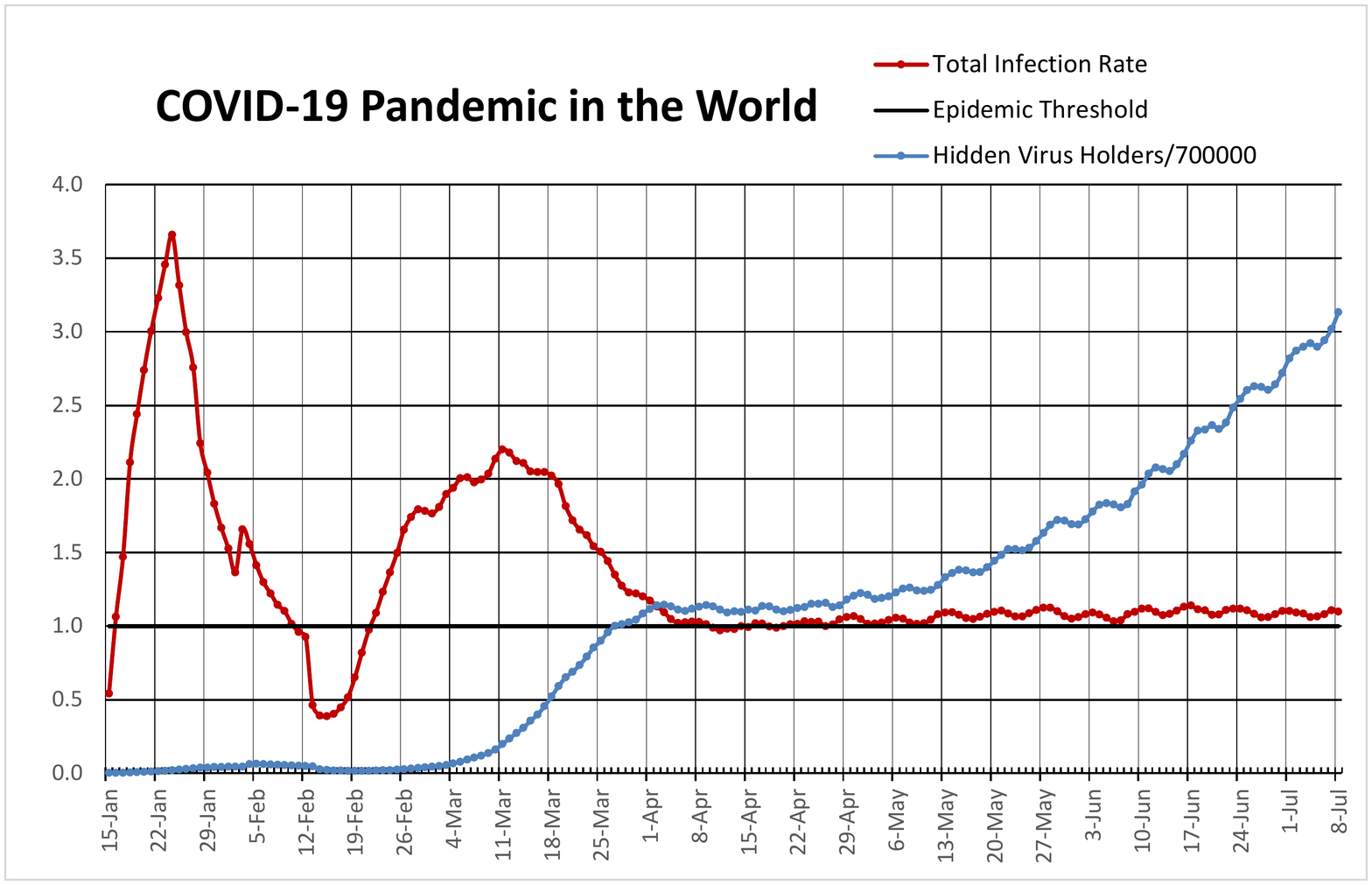} 
\caption{Development of the total infection rate $\Gamma(\cdot)$ in comparison with the estimate $H(\cdot)$ for the asymptomatic virus holders. The World, March-July 2020.}\label{fig-WrdGrow}
\end{figure}

The conclusion of our model on the overall dynamics of the pandemic in the world is represented at Figure \ref{fig-WrdGrow}. There we can observe all main phases of its development. The period January-February corresponds to the 
fight with the epidemic in China. Starting from the end of February, Europe and the US contributed in a quick growth of the total infection rate and in reducing it later by lockdown measures. After a stagnation period in April, Latin America took the lead ensuring a permanent growth of the hidden virus holders. Up to now, there is no indication of how and when it can be stopped. Most probably, the world-wide measures cannot help and every country needs to find its own way of solving the problem.

\subsection{Spain}\label{sc-Spain}

\begin{figure}[h!]
\centering
\includegraphics[scale=0.5]{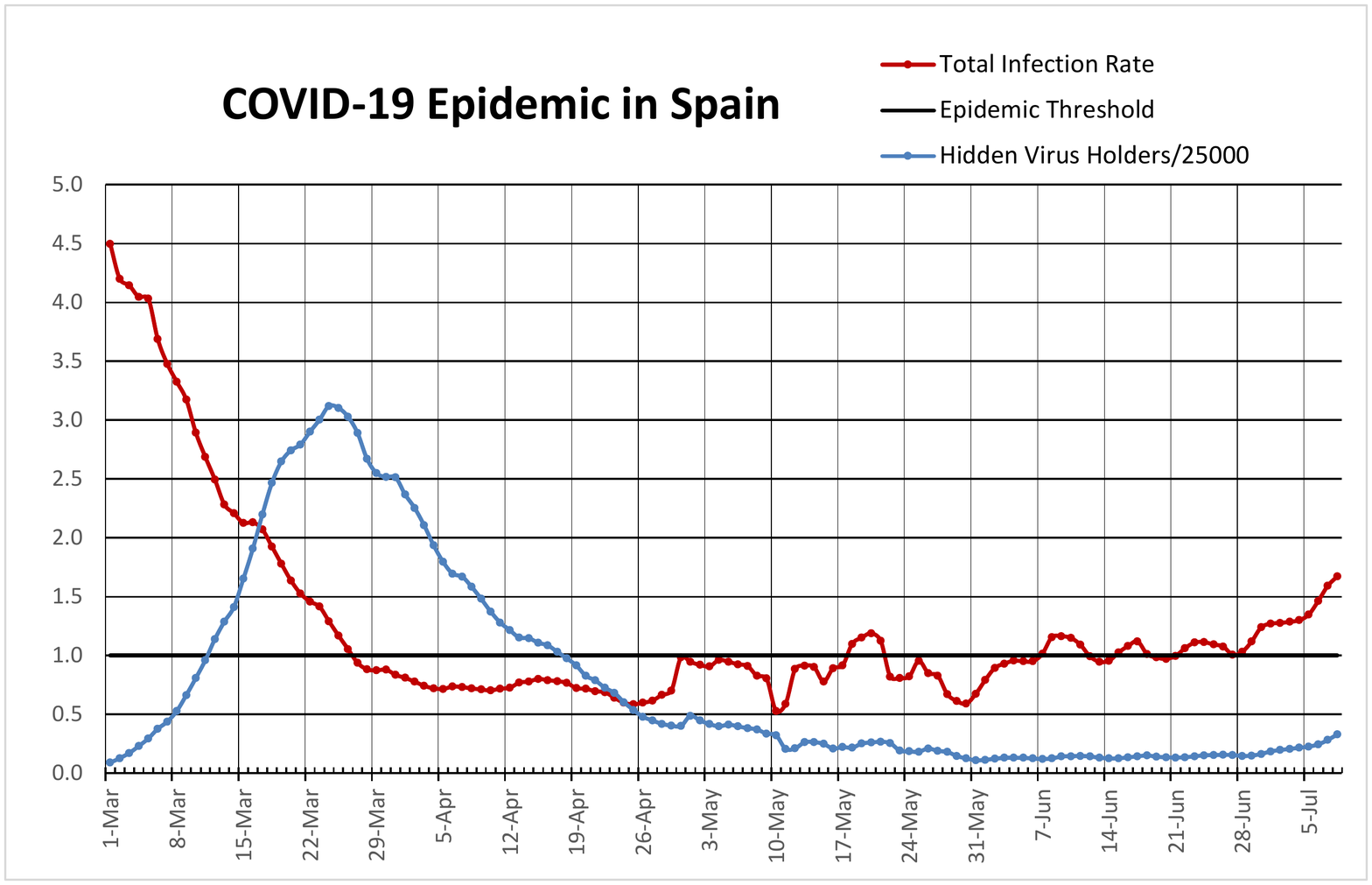} 
\caption{Development of the total infection rate $\Gamma(\cdot)$ in comparison with the estimate $H(\cdot)$ for the asymptomatic virus holders. Spain, March-July 2020.}\label{fig-SpainGrow}
\end{figure}

We finish our survey with three countries with nonstandard ways of fighting against the pandemic. 
Our first non-standard example is Spain (see Figure \ref{fig-SpainGrow}), the country which induces the second wave of the virus by attracting the foreign tourists. 

Spain was attacked by COVID-19 in the end of February. The first containment measures were taken in the beginning of March, and on March 13 the Prime Minister of Spain announced a a nationwide State of Alarm. This lockdown was very efficient. In three weeks the total infection rate was dropped below the Epidemic Threshold. 
Spain managed to avoid the stagnation phase, ensuring a permanent reduction of the asymptomatic virus holders during two months, April and May.

The essential opening started in the second half of May. On 25 May 2020, 47$\%$ of the territory of Spain was on phase 2 of lifting the lockdown restrictions. On June 8, 48$\%$ of the country was on phase 2 and 52$\%$, etc. The state of alarm expired at June 21st.
The government opened all internal borders as well as the border with France. The international flights with other European Union countries and the United Kingdom were also allowed.

As a result, starting from June 28, our estimate for the total infection rate started to grow above the Epidemic Threshold. The number of asymptomatic virus holders has already reached the level of 8000. This is still far from the historical maximum of 78000 in the end of March. However, as we know, this number can grow very quickly. 

We can expect that the available statistics underestimates the actual number of cases since starting from the beginning of July the additional pollution comes from the foreign tourists. At this moment, we have no tools for estimating its importance. 

\subsection{Sweden}\label{sc-Swed}

Let us look now at Sweden, the enfant terrible of European Health System. This is the only country in Western Europe, which completely rejected the idea of lockdown. The main motivations of this decision were as follows.
\BI
\II
Lockdown is unable to eliminate the virus completely. A small fraction of the remaining virus holders can generate next waves of the epidemic.
\II
Economical consequences of lockdown are disastrous.
\II
The virus can be killed only by the development of sufficient immunity in the society.
\II
Lockdown contradicts the Constitution of the country.
\EI
The objections to this position are also standard.
\BI
\II
This is a new virus with a high mortality rate for certain age groups.
\II
Since this is a small-scale epidemic, sufficient immunity cannot be developed.
\II
All long-term consequences of the disease are not clear yet. Hence, vaccination can be dangerous.
\EI

\begin{figure}[h!]
\centering
\includegraphics[scale=0.5]{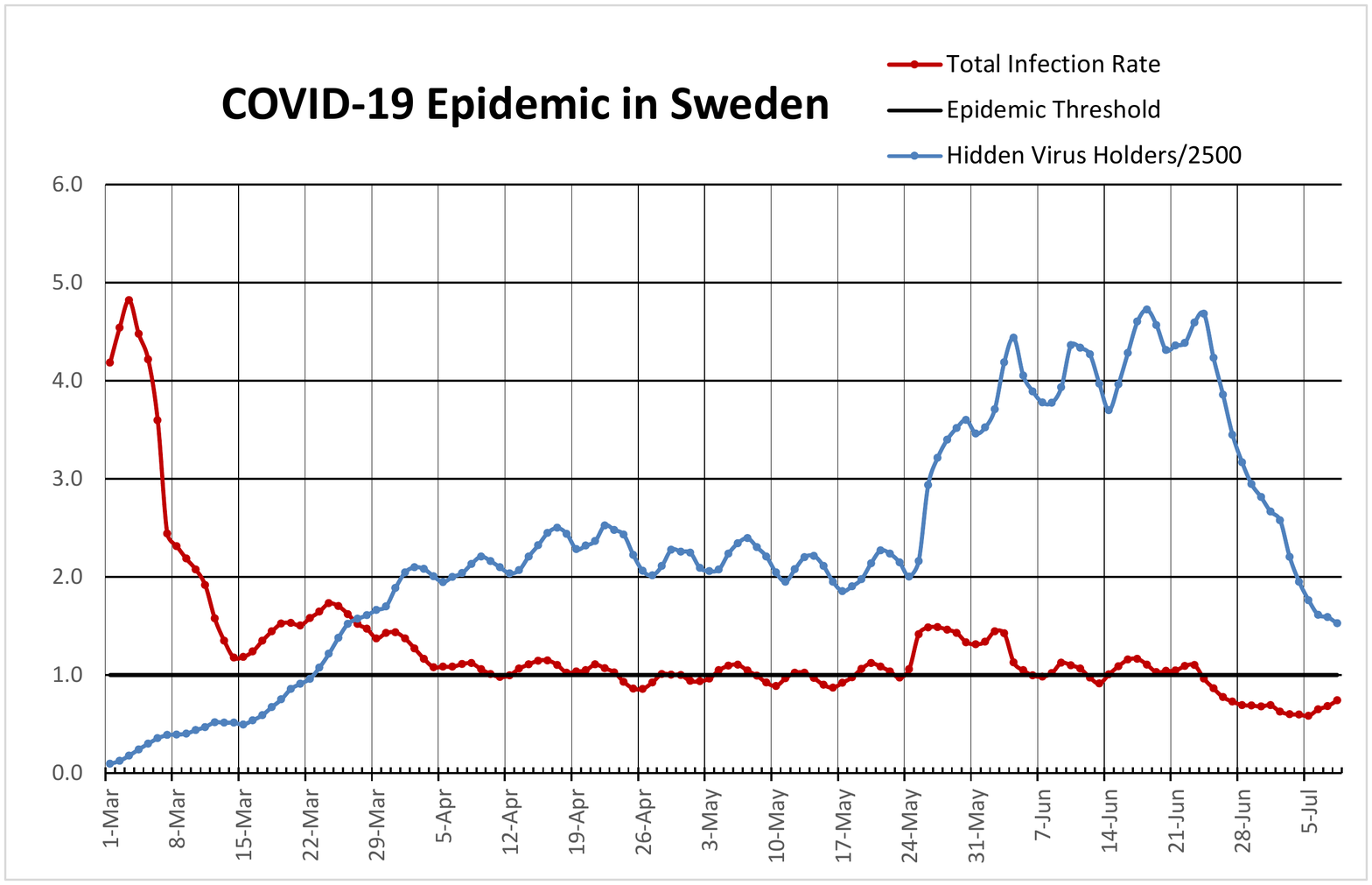} 
\caption{Development of the total infection rate $\Gamma(\cdot)$ in comparison with the estimate $H(\cdot)$ for the asymptomatic virus holders. Sweden, March-July 2020.}\label{fig-SwedGrow}
\end{figure}

In any case, Sweden did not impose a lockdown, and kept large parts of its society open. However, starting from the beginning of March, the Swedish public was expected to follow a series of non-voluntary recommendations like keeping the social distancing, restricting the size of the meetings and reducing unnecessary travel.
Let us look at the results of this strategy, represented at Figure \ref{fig-SwedGrow}.

As compared to other countries, the results are not bad. Indeed, after a fast reduction of the initial rate up to the Epidemic Threshold, Sweden entered a stagnation period of April and three first weeks of May. During this time, the number of hidden virus holders was kept on the constant level of 5000. Our model detected a sudden jump in the infection rate for two weeks of May 24 -June 7, which resulted in increasing the level of $H(\cdot)$ up to 10-11 thousands. This increase provoked a discussion in the society on the correctness of the current strategy \cite{BI} and probability of the second wave of infection \cite{Swed1}. It seems that just these fears were sufficient for strengthening the level of individual protection and dropping the total infection rate below the Epidemic Threshold. Thus, in our estimates, the number of asymptomatic virus holders in Sweden now is on the level of four thousands and it is decreasing.

\subsection{The Netherlands}\label{sc-Holl}

\begin{figure}[h!]
\centering
\includegraphics[scale=0.5]{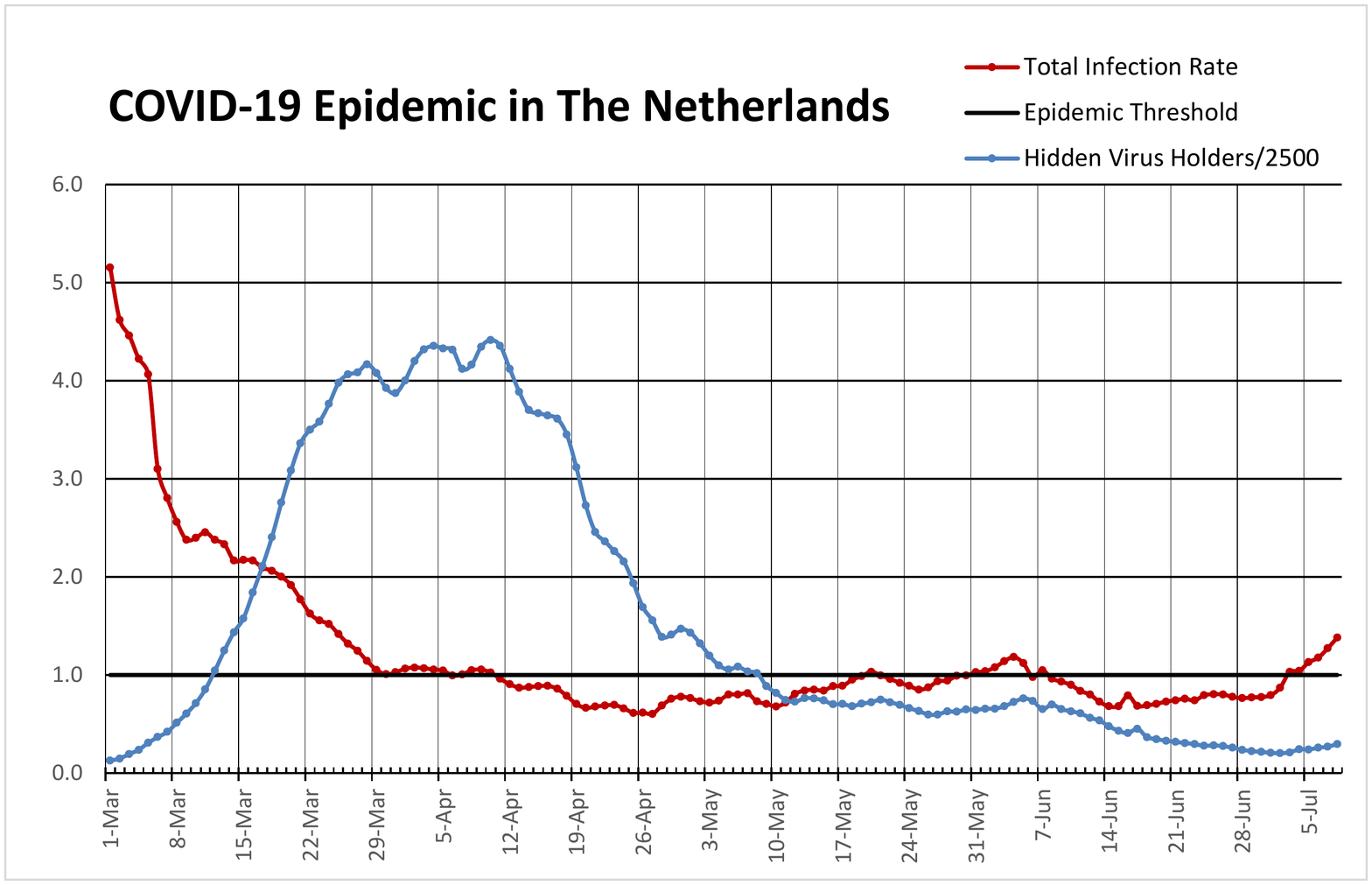} 
\caption{Development of the total infection rate $\Gamma(\cdot)$ in comparison with the estimate $H(\cdot)$ for the asymptomatic virus holders. The Netherlands, March-July 2020.}\label{fig-HollGrow}
\end{figure}

The strategy of this country can be seen as a soft variant of the Swedish plan.
On March~12,  Prime Minister {\em asked} everyone in the country to work from home as much as possible. This was the beginning of so-called 
{\em Intelligent Lockdown}. It is based on trusting the people in practicing self-distancing and different sanitizing measures. 
The whole plan relied on the public support. And Figure \ref{fig-HollGrow} confirms its success \cite{Bloom}, at least up to a certain date.

In The Netherlands, the outdoor terraces at cafes, bars, and restaurants were reopened on June 1. However, with an appropriate behavior of people, this was not dangerous. During three weeks, the total infection rate was still below the Epidemic Threshold, resulting in the minimal number of the asymptomatic virus holders of the order of 500. 

However, on July 1st, the Netherlands has lifted the travel ban for certain groups of travelers. And this was a mistake. The total infection rate started to grow immediately, pushing up the estimate $H(\cdot)$. At this moment, there is still time for necessary corrections. Otherwise, there is a real danger of the second wave.

\section{Conclusion}\label{sc-Conc}

Let us present some immediate conclusions from our survey.

{\bf 1. Scalability.} In Section \ref{sc-App}, we applied our model to countries of very different size. We looked at small Luxembourg with 626 thousands citizens, The Netherlands with 17 millions, and very big USA with 331 millions. In all cases, we get reasonable results, which are compatible with the actual history of the epidemics.

{\bf 2. Contamination delay.} Initially, the contamination delay $\Delta = 10$ was chosen by analyzing the COVID-19 epidemics in Belgium. However, the results of Section \ref{sc-App} confirm that this value is appropriate for other countries too. With this value of parameter $\Delta$, we were able to reconstruct the {\em exact dates} of important events in the past for four completely different countries: Israel, Japan, Spain, and The Netherlands.

{\bf 3. Dynamics of the new cases.} 
By many examples presented in Section \ref{sc-App}, we have shown that the conclusions based directly on the small numbers of new cases are {\em always wrong}. This mistake was repeated in Israel, Japan, Spain, and The Netherlands with very bad consequences. It seems that the only safe strategy now is to look at the level of $H(\cdot)$, trying to reduce it up to zero values, and then wait for ten days to be sure that no new cases appear.

{\bf 4. The boarders.} The analysis of the statistics shows that the most dangerous operation is the reopening of boarders. For this operation, we still have no mathematical model.

{\bf 5. Total Infection Rate.} Our new indicator $\Gamma(\cdot)$  demonstrates a good performance and remarkable accuracy. Its computation is trivial and we hope that it will be useful for further analysis of the pandemic.

{\bf 6. Daily statistics.} We can compute the indicator $\Gamma(\cdot)$ only by the daily statistics for the new cases. Its regularity and accuracy is very important for the quality of our results. Unfortunately, at some countries this service is quite unreliable.

Finally let us discuss the perspectives of eliminating the virus in the world. On the national level, the problem is easier. We have already seen the success stories of Japan, Sweden, and The Netherlands, where the number of asymptomatic virus holders was reduced up to a very small level by quite reasonable restrictions in the social behavior. The final failures in some of these countries were related to the wrong choice of the stopping moment. Now, with availability of the new indicators, it may be easier.

Unfortunately, on the international level, this problem is much more difficult and needs further intensive investigations.

\end{document}